\documentclass[12pt,english,floatfix,nofootinbib,superscriptaddress,aps,prd,preprint]{revtex4-2}
\usepackage[utf8]{inputenc}
\usepackage{float}
\usepackage{array}
\usepackage{lipsum}
\usepackage{graphicx}
\usepackage{amsmath,amsthm,amsfonts,amssymb}
\usepackage{graphicx}
\usepackage[english]{babel}
\usepackage{color}
\usepackage{physics}
\usepackage{tensor}
\usepackage{esint}
\usepackage[dvips]{epsfig}
\usepackage[dvips]{graphicx}
\usepackage{float}
\usepackage{units}
\usepackage{textcomp}
\usepackage{mathrsfs}
\usepackage{amsmath}
\usepackage[makeroom]{cancel}
\usepackage{amssymb}
\usepackage{amsbsy}
\usepackage{amsfonts}
\usepackage{amssymb,mathrsfs,xcolor}
\usepackage{esint}
\usepackage{array}
\usepackage{graphicx}
\usepackage{mathtools,slashed}
\usepackage{slashed}
\usepackage{hyperref}
\hypersetup{
    colorlinks,
    citecolor=blue,
    filecolor=green,
    linkcolor=purple,
    urlcolor=red,
}

\newcommand{\ie}{\begin{equation}}
\newcommand{\fe}{\end{equation}}
\newcommand{\bea}{\begin{eqnarray}}
\newcommand{\eea}{\end{eqnarray}}

\usepackage{hyperref}
\hypersetup{colorlinks,breaklinks,
			citecolor=[rgb]{0,0.0,1.0},
            urlcolor=[rgb]{0.0,0.0,1.0},
            linkcolor=[rgb]{0,0.5,0.9}}

\begin{document}

\title{Nonlocal spinor superfield theory}

\author{F. S. Gama}
\email{fgama@unifap.br}
\affiliation{Departamento de Ci\^{e}ncias Exatas e Tecnol\'{o}gicas, Universidade Federal do Amap\'{a}, 68903-419, Macap\'{a}, Amap\'{a}, Brazil}

\author{J. R. Nascimento}
\email{jroberto@fisica.ufpb.br}
\affiliation{Departamento de Física, Universidade Federal da Paraíba, Caixa Postal 5008, 58051-970, João Pessoa, Paraíba,  Brazil} 

 \author{Gonzalo J. Olmo}
\email[]{gonzalo.olmo@uv.es}
\affiliation{Instituto de Física Corpuscular (IFIC), CSIC‐Universitat de València, Spain}
\affiliation{Universidade Federal do Cear\'a (UFC), Departamento de F\'isica,\\ Campus do Pici, Fortaleza - CE, C.P. 6030, 60455-760 - Brazil.}	

\author{A. Yu. Petrov}
\email{petrov@fisica.ufpb.br}
\affiliation{Departamento de Física, Universidade Federal da Paraíba, Caixa Postal 5008, 58051-970, João Pessoa, Paraíba,  Brazil} 

\author{P. J. Porfírio}
\email{pporfirio@fisica.ufpb.br}
\affiliation{Departamento de Física, Universidade Federal da Paraíba, Caixa Postal 5008, 58051-970, João Pessoa, Paraíba,  Brazil}


\date{\today}

\begin{abstract}

In this work, we propose a new three-dimensional nonlocal spinor superfield model. This theory is constructed by introducing form factors in the local spinor superfield action. Then, we couple it minimally to a scalar superfield, 
for which we calculate the one-loop effective potential as a first constructive example of perturbative calculations in this new theory. 

\end{abstract}

\maketitle
\newpage

\section{INTRODUCTION}

The initial motivations to consider nonlocal field theories stemmed from the need for an adequate description of finite-size objects \cite{Efimov}, motivated by elementary particle physics, and also because of the need for a consistent quantum theory of gravity. In this sense, it is well known that the standard formulation of Einstein's gravity is non-renormalizable, and a natural way to improve its renormalizability properties is to add higher-derivative terms, as they can strongly improve the convergence of loop integrals. However, it is well known that adding higher derivatives may lead to the appearance of ghosts (see, for example, \cite{Hawking:2001yt} and references therein). In many relevant cases, nonlocal field theories are known to be ghost-free and even display super-renormalizability or finiteness, thus becoming sound candidates for a consistent quantum formulation of gravity (see for example \cite{Modesto:2011kw,Modesto:2016ofr} and references therein).

 Nonlocal extensions of non-gravitational field theories provide convenient laboratories for investigating the consequences of nonlocality in a controlled setting. In this context, it is worth mentioning the study of loop corrections in a nonlocal scalar field theory model \cite{Briscese:2015zfa}, as well as the superfield generalization of this model \cite{BezerradeMello:2016bjn}, formulated in four-dimensional spacetime with chiral and antichiral superfields whose components include scalar fields, see for example \cite{GGRS}. A further extension of this theory can be obtained by minimal coupling to a gauge superfield \cite{Gama:2017ets,Gama:2020pte}. The non-locality is consistently implemented in such models via the inclusion of form factors of the type $f(\Box)$ in their actions. Such constructions, however, are not the most general possibility, as they still leave room for generalizations of the form-factor approach based on the Dirac operator $f(\slashed{\partial})$, as recently shown in   \cite{Nascimento:2025ngc}.  Nonetheless, in the context of  $f(\Box)$ nonlocality, it becomes necessary to find if superfield analogues of such theories can be consistently built and, second, if models describing the (three-dimensional) spinor supermultiplet discussed in \cite{GGRS,dualSUSY} are also possible. These are the main goals pursued in the present work. For concreteness and simplicity, here we focus our attention on a simpler three-dimensional model. This is advantageous because the standard four-component spinors in a four-dimensional spacetime correspond to a reducible representation of the Lorentz group, which adds unnecessary complications in the calculus. We plan to treat the four-dimensional case elsewhere. Thus, in this work, we construct the three-dimensional free nonlocal spinor superfield action based on form-factors of the form $f(\Box)$, then couple it to the external scalar superfield, and calculate the resulting effective potential.

The structure of the paper is as follows. In Section 2, we write down the classical action of the theory. In Section 3, we obtain the effective potential of one-loop, using functional methods. Section 4 is our Summary.

\section{Formulation of the theory}

Before defining our nonlocal theory, we begin by writing down the local three-dimensional model of the spinor superfield proposed in \cite{dualSUSY}:
\begin{equation}
	\label{dirac}
	S_\Psi=\int d^5z \bar{\Psi}^{\alpha}\left(i{\partial_{\alpha}}^{\beta}+M_\Psi{\delta_{\alpha}}^{\beta}\right)\Psi_{\beta},
\end{equation}
where $\Psi^{\alpha}$ denotes the spinor superfield and $\bar{\Psi}^{\alpha}$ its conjugate. For a detailed discussion of the component field content of (\ref{dirac}), we refer the reader to \cite{dualSUSY}. The model above can be treated as a $\mathcal{N}=1$, $d=3$ analog of the standard Dirac action defined in Minkowski spacetime.

{ Here we note that this action, rewritten within the component formalism, does not contain the spinor action studied in \cite{Nascimento:2025ngc}, which can arise from some action of the scalar multiplet described by the scalar superfield $\Phi$. Indeed, it is known \cite{GGRS} that the spinor field is defined within the scalar multiplet as $\psi_{\alpha}=D_{\alpha}\Phi|$, where the vertical bar means that after taking the derivative, the Grassmann coordinates are put to zero. As is well known (cf. \cite{GGRS}), one has
\begin{equation}
\frac{1}{2}\int d^5z \Phi(D^2-m)\Phi=\frac{1}{2}\int d^3x \psi^{\alpha}(i\partial_{\alpha}^{\phantom{\alpha}\beta}-m\delta_{\alpha}^{\beta})\psi_{\beta}+\ldots,
\end{equation}
where dots are for other components of $\Phi$. At the same time, the action (\ref{dirac}) within the component formulation involves not one but two independent complex spinor fields together with the conjugated ones, see \cite{dualSUSY}, hence its component structure strongly differs from \cite{Nascimento:2025ngc}. Nevertheless, the action (\ref{dirac}) is especially interesting for us because it displays a nontrivial matrix structure of the operator describing the quadratic action, in the same spirit as the one discussed in \cite{Nascimento:2025ngc}. Therefore (\ref{dirac}) can be seen as a superfield analogue of the theory considered there.}

Note that (\ref{dirac}) is invariant under global $U(1)$ transformations, which allows for a minimal coupling to the gauge superfield $A_\alpha$. However, since the main objective of this work is to determine the superfield effective potential (SEP), which by definition depends only on scalar superfields, we shall instead consider the interaction of the spinor superfield with a dynamical real scalar superfield, whose action is conventionally defined by \cite{GGRS}
\begin{equation}
	\label{scalar}
	S_\Phi=\frac{1}{2}\int d^5z \Phi\left(D^2+M_\Phi\right)\Phi.
\end{equation}

We now extend $S_\Psi + S_\Phi$ to include interactions between $\Psi^\alpha$ and $\Phi$. Among the several possibilities that could in principle be considered, we restrict ourselves to adding interaction terms that preserve the renormalizability of the complete theory, namely those associated with coupling constants of non-negative mass dimension. In addition, we disregard terms of cubic and higher order in $\Psi^\alpha$, since such contributions do not affect the SEP at the one-loop level. Under these conditions, we assume that the interaction between $\Psi^\alpha$ and $\Phi$ is described by
\begin{equation}
	\label{interaction}
	S_{INT}=\int d^5z\left[X(\Phi)+\frac{\lambda}{2}\Phi^2\left(D^\alpha\Psi_\alpha+D^\alpha\bar \Psi_\alpha\right)+Y(\Phi)\bar \Psi^\alpha \Psi_\alpha+Z(\Phi)\left(\Psi^2+\bar \Psi^2\right)\right],
\end{equation}
where
\begin{equation}
	\label{potentials}
	X(\Phi)=\frac{a_3}{3!}\Phi^3+\frac{a_4}{4!}\Phi^4 \ ; \ Y(\Phi)=\eta_1\Phi+\frac{\eta_2}{2}\Phi^2 \ ; \ Z(\Phi)=\mu_1\Phi+\frac{\mu_2}{2}\Phi^2.
\end{equation}

By adding Eqs. (\ref{dirac}), (\ref{scalar}), and (\ref{interaction}), we obtain the following complete local superfield theory:
\begin{equation}
	\label{local}
	\begin{split}
		S_{local}=&\int d^5z\left[ \bar{\Psi}^{\alpha}\left(i{\partial_{\alpha}}^{\beta}+M_\Psi{\delta_{\alpha}}^{\beta}\right)\Psi_{\beta}+\Phi\left(D^2+M_\Phi\right)\Phi\right]\\
		&+\int d^5z\left[X(\Phi)+\frac{\lambda}{2}\Phi^2\left(D^\alpha\Psi_\alpha+D^\alpha\bar \Psi_\alpha\right)+Y(\Phi)\bar \Psi^\alpha \Psi_\alpha+Z(\Phi)\left(\Psi^2+\bar \Psi^2\right)\right].
	\end{split}
\end{equation}
The next step is to construct a higher-derivative and further nonlocal generalization of this action. This can be achieved by introducing the operators $f_1(\Box)$, $f_2(\Box)$ and $g_1(\Box)$, $g_2(\Box)$ into the quadratic terms of $\Psi^\alpha$ and $\Phi$, respectively. In this way, we obtain:
\begin{equation}
	\label{nonlocal}
	\begin{split}
		S=&\int d^5z\left[ \bar{\Psi}^{\alpha}\left(if_1(\Box){\partial_{\alpha}}^{\beta}+M_\Psi{f_2(\Box)\delta_{\alpha}}^{\beta}\right)\Psi_{\beta}+\Phi\left(g_1(\Box)D^2+M_\Phi g_2(\Box)\right)\Phi\right]\\
		&+\int d^5z\left[X(\Phi)+\frac{\lambda}{2}\Phi^2\left(D^\alpha\Psi_\alpha+D^\alpha\bar \Psi_\alpha\right)+Y(\Phi)\bar \Psi^\alpha \Psi_\alpha+Z(\Phi)\left(\Psi^2+\bar \Psi^2\right)\right].
	\end{split}
\end{equation}
It is well known that if the dimensionless operators $f_1(\Box)$, $f_2(\Box)$, $g_1(\Box)$, and $g_2(\Box)$ are taken to be polynomial functions of the d’Alembertian operator, the resulting theory generally contains additional unphysical degrees of freedom (ghosts) \cite{Ghosts}. To circumvent this problem, we assume that these functions are nonlocal operators chosen in such a way that no ghosts are introduced \cite{NL}. Moreover, they must reduce to the identity operator in an appropriate limit, ensuring recovery of the local theory (\ref{local}). For instance, one of the simplest choices that meets both requirements is
\begin{equation}
	f_1(\Box)=f_2(\Box)=g_1(\Box)=g_2(\Box)=\exp\left(-\frac{\Box}{\Lambda^2}\right).
\end{equation}
In this case, no ghostlike degrees of freedom are introduced, since the exponential operator $\exp\left(-\frac{\Box}{\Lambda^2}\right)$ is an entire function without zeros and smoothly approaches the identity operator in the limit $\Lambda \to \infty$. Moreover, the resulting theory exhibits an improved ultraviolet behavior relative to the local action (\ref{local}). This choice has been widely used in earlier studies, see e.g. \cite{BezerradeMello:2016bjn}.

\section{One-loop Effective Potential}

In this section, we compute the one-loop SEP associated with the theory defined in (\ref{nonlocal}). To this end, we employ the background field method~\cite{DeWitt,BOS}. According to this procedure, the superfields in (\ref{nonlocal}) are linearly decomposed into sums of background components and quantum fluctuations as follows:
\begin{equation}
	\Psi^\alpha\rightarrow\Psi^\alpha+\psi^\alpha \ ; \ \bar \Psi^\alpha\rightarrow\bar \Psi^\alpha+\bar \psi^\alpha \ ; \ \Phi\rightarrow\Phi+\phi.
\end{equation}
However, since by definition the SEP depends only on the background scalar superfield $\Phi$~\cite{FGLNPS}, we set the background spinor superfields $\Psi^\alpha$ and $\bar \Psi^\alpha$, as well as the covariant derivatives of $\Phi$, to be zero. Furthermore, because we are concerned solely with the one-loop contribution, it is sufficient to expand the action (\ref{nonlocal}) and retain only the terms quadratic in the quantum superfields. After performing these straightforward algebraic manipulations, we obtain:
\begin{equation}
	\label{quadratic}
	S_2=\int d^5z\left\{\frac{1}{2}\begin{pmatrix}
		\psi^\alpha & \bar{\psi}^\alpha
	\end{pmatrix}{\mathcal{O}_\alpha}^\beta\begin{pmatrix}
	\psi_\beta\\ \bar \psi_\beta
	\end{pmatrix}+\begin{pmatrix}
	\psi^\alpha & \bar{\psi}^\alpha
	\end{pmatrix}\mathcal{F}_\alpha+\frac{1}{2}\phi\left[g_1(\Box)D^2+\widetilde M_\Phi(\Box)\right]\phi\right\},
\end{equation}
where 
\begin{equation}
	\widetilde M_\Psi(\Box)=M_\Psi f_2(\Box)+Y(\Phi) \ ; \ \widetilde M_\Phi(\Box)=M_\Phi g_2(\Box)+X^{\prime\prime}(\Phi)
\end{equation}
and
\begin{align}
	\label{operator}
	{\mathcal{O}_\alpha}^\beta&=\begin{pmatrix}
		Z(\Phi){\delta_\alpha}^\beta & f_1(\Box)i{\partial_\alpha}^\beta+\widetilde M_\Psi(\Box){\delta_\alpha}^\beta\\
		f_1(\Box)i{\partial_\alpha}^\beta+\widetilde M_\Psi(\Box){\delta_\alpha}^\beta & Z(\Phi){\delta_\alpha}^\beta
	\end{pmatrix};\\
	\mathcal{F}_\alpha&=\begin{pmatrix}
		-\lambda\Phi D_\alpha\phi \\
		-\lambda\Phi D_\alpha\phi
	\end{pmatrix}.
\end{align}
At this stage, the calculations become increasingly cumbersome, which motivates us to introduce a simplifying assumption. From this point onward, we consider the particular case of our theory assuming that
\begin{equation}
	\label{condition}
	Z(\Phi)=0,
\end{equation}
a condition that renders all diagonal elements of ${\mathcal{O}_\alpha}^{\ \beta}$ equal to zero.

To proceed further, we must eliminate the undesired mixing terms proportional to $\psi^\alpha D_\alpha\phi$ and $\bar{\psi}^\alpha D_\alpha\phi$ that appear in (\ref{quadratic}). This can be accomplished by performing the following nonlocal change of variables in the path integral \cite{NL-transf}:
\begin{equation}
	\label{transf}
	\begin{pmatrix}
		\psi^\alpha\\ \bar \psi_\alpha
	\end{pmatrix}\longrightarrow\begin{pmatrix}
		\psi^\alpha\\ \bar \psi_\alpha
	\end{pmatrix}-{(\mathcal{O}^{-1})_\alpha}^{\beta}\mathcal{F}_\beta,
\end{equation}
where
\begin{equation}
	{(\mathcal{O}^{-1})_\alpha}^\beta=\frac{1}{\Box f_1^2(\Box)-\widetilde M_\Psi^2(\Box)}\begin{pmatrix}
	0 & f_1(\Box)i{\partial_\alpha}^\beta-\widetilde M_\Psi(\Box){\delta_\alpha}^\beta\\
	f_1(\Box)i{\partial_\alpha}^\beta-\widetilde M_\Psi(\Box){\delta_\alpha}^\beta & 0
	\end{pmatrix}.
\end{equation}
Since the transformations (\ref{transf}) correspond to constant shifts in the integration variables, the functional integration measures remain unchanged.

Substituting (\ref{transf}) into (\ref{quadratic}) and making use of the identities
\begin{align}
	\int d^5z\left((\mathcal{O}^{-1})^{\alpha\gamma}\mathcal{F}_\gamma\right)^T{\mathcal{O}_\alpha}^\beta\begin{pmatrix}
		\psi_\beta\\ \bar\psi_\beta
	\end{pmatrix}&=\int d^5z\begin{pmatrix}
	\psi^\alpha & \bar\psi^\alpha
	\end{pmatrix}\mathcal{F}_\alpha;\\
	-\frac{1}{2}\int d^5z\left((\mathcal{O}^{-1})^{\alpha\beta}\mathcal{F}_\beta\right)^T\mathcal{F}_\alpha&=2\lambda^2\int d^5z\Phi^2\phi\frac{\Box f_1(\Box)-\widetilde M_\Psi(\Box)D^2}{\Box f_1^2(\Box)-\widetilde M_\Psi^2(\Box)}\phi,
\end{align}
we obtain, after straightforward algebraic manipulations,
\begin{equation}
	\begin{split}
		\label{quad_trans}
		S_2&=\int d^5z\Bigg\{\frac{1}{2}\begin{pmatrix}
		\psi^\alpha & \bar{\psi}^\alpha
		\end{pmatrix}{\mathcal{O}_\alpha}^\beta\begin{pmatrix}
		\psi_\beta\\ \bar \psi_\beta
		\end{pmatrix}\\
		&+\frac{1}{2}\phi\left[g_1(\Box)D^2+\widetilde M_\Phi(\Box)+4\lambda^2\Phi^2\frac{\Box f_1(\Box)-\widetilde M_\Psi(\Box)D^2}{\Box f_1^2(\Box)-\widetilde M_\Psi^2(\Box)}\right]\phi\Bigg\}.
	\end{split}
\end{equation}
By integrating out the quantum superfields in (\ref{quad_trans}), we arrive at two contributions to the Euclidean one-loop effective action:
\begin{equation}
	\label{EA_1}
	\Gamma^{(1)}=\frac{1}{2}\operatorname{Tr}\ln{\mathcal{O}_\alpha}^\beta-\frac{1}{2}\operatorname{Tr}\ln\left[g_1(\Box)D^2+\widetilde M_\Phi(\Box)+4\lambda^2\Phi^2\frac{\Box f_1(\Box)-\widetilde M_\Psi(\Box)D^2}{\Box f_1^2(\Box)-\widetilde M_\Psi^2(\Box)}\right].
\end{equation}
Although the operator ${\mathcal{O}_\alpha}^\beta$ depends on the background superfield [see Eq. (\ref{operator})], it does not contain spinor covariant derivatives. Therefore, the first trace in (\ref{EA_1}) vanishes due to the properties of the Grassmann delta function.
On the other hand, the second trace can be conveniently decomposed into three parts:
\begin{equation}
	\begin{split}
		\label{EA_2}
		\Gamma^{(1)}=&-\frac{1}{2}\operatorname{Tr}\ln\left[g_1(\Box)D^2\right]-\frac{1}{2}\operatorname{Tr}\ln\left[1+\frac{\widetilde M_\Phi(\Box)}{\Box g_1(\Box)}D^2\right]\\
		&-\frac{1}{2}\operatorname{Tr}\ln\left[1+\frac{4\lambda^2\Phi^2\left(\Box f_1(\Box)g_1(\Box)+\widetilde M_\Psi(\Box)\widetilde M_\Phi(\Box)\right)}{\mathcal{P}_\Psi(\Box)\mathcal{P}_\Phi(\Box)-4\lambda^2\Phi^2\Box\left( f_1(\Box)\widetilde M_\Phi(\Box)+ g_1(\Box)\widetilde M_\Psi(\Box)\right)}D^2\right],
	\end{split}
\end{equation}
where
\begin{equation}
	\mathcal{P}_\Psi(\Box)=\Box f_1^2(\Box)-\widetilde M_\Psi^2(\Box) \ ; \ \mathcal{P}_\Phi(\Box)=\Box g_1^2(\Box)-\widetilde M_\Phi^2(\Box).
\end{equation}
The first trace is independent of the background superfield and may therefore be discarded since the Grassmann integral of a constant vanishes. The second and third traces, in contrast, can be evaluated by invoking the definition of the functional trace \cite{BK}, that is,
\begin{equation}
	\begin{split}
		\operatorname{Tr}\ln\left[1+A(\Box)D^2\right]=&\int d^5z\int d^5z^\prime\delta^5(z^\prime-z)\ln\left[1+A(\Box)D^2\right]\delta^5(z-z^\prime)\\
		=&\int d^5z\int d^5z^\prime\delta^5(z^\prime-z)\sum_{n=1}^{\infty}(-1)^{n+1}\frac{A^n(\Box)}{n}(D^2)^n\delta^5(z-z^\prime).
	\end{split}
\end{equation}
Using $(D^2)^2=\Box$ together with the standard identities
\begin{equation}
	\delta^2(\theta^\prime-\theta)\delta^5(z-z^\prime)=0 \ ; \ \delta^2(\theta^\prime-\theta)D^2\delta^5(z-z^\prime)=\delta^5(z-z^\prime),
\end{equation}
we conclude that only the terms with odd $n$ contribute. Hence,
\begin{equation}
	\begin{split}
		\operatorname{Tr}\ln\left[1+A(\Box)D^2\right]=&\int d^5z\int d^5z^\prime\delta^5(z^\prime-z)\sum_{k=0}^{\infty}\frac{A^{2k+1}(\Box)}{2k+1}(D^2)^{2k+1}\delta^5(z-z^\prime)\\
		=&\int d^5z\int d^3x^\prime\delta^3(x^\prime-x)\frac{1}{\sqrt{\Box}}\operatorname{arctanh}\left(A(\Box)\sqrt{\Box}\right)\delta^3(x-x^\prime).
	\end{split}
\end{equation}
Going to momentum space, and using the identity $\arctan(x)=-i\operatorname{arctanh}(ix)$, we obtain
\begin{equation}
		\operatorname{Tr}\ln\left[1+A(\Box)D^2\right]=\int d^5z\int \frac{d^3p}{(2\pi)^3}\frac{1}{\abs{p}}\arctan\left(\abs{p} A(-p^2)\right).
\end{equation}
Substituting this expression into (\ref{EA_2}), and using the
relation $\Gamma^{(1)}=\int d^5zK^{(1)}$, we finally obtain the one-loop correction to the SEP:
\begin{equation}
	\label{main_result}
	\begin{split}
		&K^{(1)}=\frac{1}{2}\int \frac{d^3p}{(2\pi)^3}\frac{1}{\abs{p}}\arctan\left(\frac{\widetilde M_\Phi(-p^2)}{\abs{p} g_1(-p^2)}\right)-\frac{1}{2}\int \frac{d^3p}{(2\pi)^3}\frac{1}{\abs{p}}\\
		&\times\arctan\left( \frac{4\lambda^2\Phi^2\abs{p}\left(\widetilde M_\Psi(-p^2)\widetilde M_\Phi(-p^2)-p^2 f_1(-p^2)g_1(-p^2)\right)}{\mathcal{P}_\Psi(-p^2)\mathcal{P}_\Phi(-p^2)+4\lambda^2\Phi^2p^2\left( f_1(-p^2)\widetilde M_\Phi(-p^2)+ g_1(-p^2)\widetilde M_\Psi(-p^2)\right)}\right),
	\end{split}
\end{equation}
valid whenever condition (\ref{condition}) is satisfied.

The SEP presented in (\ref{main_result}) constitutes our main result. In order to obtain an explicit expression for $K^{(1)}$, one must specify the nonlocal operators $f_{1}(\Box)$, $f_{2}(\Box)$, $g_{1}(\Box)$, and $g_{2}(\Box)$, and subsequently evaluate the corresponding momentum integrals. However, this proves to be a difficult task even for the simplest choices of nonlocal operators, since the resulting integrals cannot be exactly solved. For this reason, we now introduce a set of simplifying assumptions that will allow us to derive explicit expressions for $K^{(1)}$ in two representative situations: first, a simplified situation, namely, a local model without higher derivatives and second, a nonlocal model (with infinitely many higher derivative terms). For the first choice, we will obtain an exact expression in terms of elementary functions, whereas for the second one, we will get an approximate expression for $K^{(1)}$.

For the first model, let us assume that
\begin{equation}
	M_\Psi=Y(\Phi)=0.
\end{equation}
In addition, we take all differential operators to coincide with the identity, namely,
\begin{equation}
	f_{1}(\Box)=f_{2}(\Box)=g_{1}(\Box)=g_{2}(\Box)=1.
\end{equation}
Under these assumptions, the general expression (\ref{main_result}) becomes considerably simpler and reduces to
\begin{equation}
	\label{integrals_local}
	\begin{split}
		K^{(1)}_L&=\frac{1}{2}\int \frac{d^3p}{(2\pi)^3}\frac{1}{\abs{p}}\arctan\left(\frac{ M_\Phi+X^{\prime\prime}}{\abs{p}}\right)\\
		&+\frac{1}{2}\int \frac{d^3p}{(2\pi)^3}\frac{1}{\abs{p}}\arctan\left( \frac{4\lambda^2\Phi^2\abs{p}}{p^2+\left(M_\Phi+X^{\prime\prime}\right)^2+4\lambda^2\Phi^2\left( M_\Phi+X^{\prime\prime}\right)}\right).
	\end{split}
\end{equation}
The first integral is well known, and one can show that it yields 
$-\frac{1}{16\pi}\left(M_\Phi+X^{\prime\prime}\right)^2$. To compute the second integral, we observe that it can be rewritten in the form of a standard one-loop Feynman integral. Indeed,
\begin{equation}
	\label{trick}
	\begin{split}
		&\int \frac{d^3p}{(2\pi)^3}\frac{1}{\abs{p}}\arctan\left( \frac{b\abs{p}}{p^2+a^2+ba}\right)\\
		&=\int \frac{d^3p}{(2\pi)^3}\int_0^1 d\gamma\frac{\partial}{\partial\gamma}\left[\frac{1}{\abs{p}}\arctan\left( \frac{\gamma b\abs{p}}{p^2+a^2+\gamma ba}\right)\right]=\int_0^1 d\gamma\int \frac{d^3p}{(2\pi)^3}\frac{b}{p^2+\left(a+\gamma b\right)^2}\\
		&=-\frac{b}{4\pi}\left(a+\frac{b}{2}\right).
	\end{split}
\end{equation}
Therefore, using the identity above in the expression (\ref{integrals_local}), we obtain the exact one-loop correction to the SEP for the local model under consideration:
\begin{equation}
	\label{local_final}
	K^{(1)}_L=-\frac{1}{16\pi}\left(M_\Phi+X^{\prime\prime}(\Phi)\right)^2-\frac{\lambda^2\Phi^2}{2\pi}\left(M_\Phi+X^{\prime\prime}(\Phi)+2\lambda^2\Phi^2\right).
\end{equation}
This SEP is UV finite and, according to Eq. (\ref{potentials}), its functional form reduces to a polynomial. This outcome is not surprising: in three-dimensional theories, within the dimensional regularization framework which we use, UV divergences generally appear only from the two-loop level onward, and it is precisely at that order that logarithmic contributions to the SEP typically emerge.

Now, for our second model, we assume that
\begin{equation}
	M_\Phi=M_\Psi=X(\Phi)=Y(\Phi)=0.
\end{equation}
Under these conditions, it is sufficient to introduce only the following nonlocal differential operators:
\begin{equation}
	\label{nonlocal_choice}
	f_{1}(\Box)=g_{1}(\Box)=\exp\left(-\frac{\Box}{\Lambda^2}\right),
\end{equation}
where $\Lambda$ denotes the characteristic scale at which nonlocal effects become significant.

With these choices, the expression (\ref{main_result}) reduces to a compact form
\begin{equation}
	K^{(1)}_{NL}=\frac{1}{2}\int \frac{d^3p}{(2\pi)^3}\frac{1}{\abs{p}}\arctan\left(\frac{4\lambda^2\Phi^2e^{-2\frac{p^2}{\Lambda^2}}}{\abs{p}}\right)
\end{equation}
Despite its apparently simple structure, the integral above cannot be evaluated in a closed form, in contrast with the exact expression found for the local case in (\ref{local_final}). For this reason, we proceed by obtaining an approximate expression for $K^{(1)}_{NL}$ in the regime where the nonlocality scale $\Lambda$ is taken to be very large, so that the nonlocal contributions can be treated perturbatively. 

Consider the asymptotic power series expansion of $K^{(1)}_{NL}$ in the limit $\Lambda\to\infty$, given by
\begin{equation}
	\label{f_int}
	\begin{split}
		&K^{(1)}_{NL}=\int \frac{d^3p}{(2\pi)^3}\Bigg[\frac{1}{2\abs{p}}\arctan\left(\frac{4\lambda^2\Phi^2}{\abs{p}}\right)-\frac{4\lambda^2\Phi^2}{\Lambda^2}\frac{p^2}{p^2+\left(4\lambda^2\Phi^2\right)^2}\\
		&+\frac{4\lambda^2\Phi^2}{\Lambda^4}\frac{p^4\left(p^2-\left(4\lambda^2\Phi^2\right)^2\right)}{\left(p^2+\left(4\lambda^2\Phi^2\right)^2\right)^2}-\frac{8\lambda^2\Phi^2}{3\Lambda^6}\frac{p^6\left(p^4-6\left(4\lambda^2\Phi^2\right)^2 p^2+\left(4\lambda^2\Phi^2\right)^4\right)}{\left(p^2+\left(4\lambda^2\Phi^2\right)^2\right)^3}+\cdots\Bigg].
	\end{split}
\end{equation}
Fortunately, all the integrals appearing in this expansion are ultraviolet finite in three dimensions. This property would not hold had the theory been defined in four dimensions, where the strategy of expansion by regions would be required to properly handle divergent integrals \cite{ER}. The first term in (\ref{f_int}) can be evaluated employing the same technique used in (\ref{trick}), while the remaining integrals follow directly from standard one–loop Feynman integral formulae.

Collecting all contributions, we obtain the following approximate expression for the one-loop correction to the SEP of the nonlocal model:
\begin{equation}
	\label{nonlocal_final}
	K^{(1)}_{NL}\approx-\frac{\left(\lambda\Phi\right)^4}{\pi}\left(1+\frac{4}{\Lambda^2}\left(2\lambda\Phi\right)^4+\frac{24}{\Lambda^4}\left(2\lambda\Phi\right)^8+\frac{512}{3\Lambda^6}\left(2\lambda\Phi\right)^{12}\right)+\ldots.
\end{equation}
This expression is fully consistent with the local limit: as $\Lambda\to\infty$, the nonlocal operators in (\ref{nonlocal_choice}) reduce to the identity, and (\ref{nonlocal_final}) smoothly reproduces the local result (\ref{local_final}). Furthermore, the functional form of (\ref{nonlocal_final}) involves only positive powers of the superfield $\Phi$, which reflects the ultraviolet finiteness of all the integrals in (\ref{f_int}), in close analogy with the structure encountered in the local case.

\section{Summary}

We formulated a new three-dimensional nonlocal model for a spinor superfield. This theory can be treated as a certain analogue of the model defined in \cite{Nascimento:2025ngc}, since, similarly to that model, the nonlocal action of our theory involves both the $\partial_{\alpha\beta}$ operator, which represents the $\slashed{\partial}$ one with explicitly given indices, and the unit matrix. For this theory, we introduced an analogue of the Yukawa coupling to a scalar superfield and, in the resulting theory, we integrated out the quantum spinor superfields and calculated the one-loop effective potential for the scalar, which is explicitly finite as expected. We found that the effects of the nonlocality in the one-loop SEP are proportional to $\frac{1}{\Lambda^2}$, becoming relevant at high-energy scales and highly suppressed at low-energy ones.

We expect that our model can be used as a prototype for studying the more involved, in particular, phenomenologically interesting nonlocal theories involving scalar and spinor superfields. Natural extensions of our study could consist of generalizing it to four dimensions and including gauge superfields. We plan to perform these studies in forthcoming papers.

{\bf Acknowledgments.} 
 This work was supported by Conselho Nacional de Desenvolvimento Cient\'{\i}fico e Tecnol\'{o}gico (CNPq), Paraiba State Research Foundation (FAPESQ-PB), and the Spanish National Grants PID2023-149560NB-C21 and the Severo Ochoa Excellence Grant CEX2023-001292-S, funded by MICIU/AEI/10.13039/501100011033 (“ERDF A way of making Europe”, “PGC Generacion de Conocimiento”) and FEDER, UE. PJP would like to thank the Brazilian agency CNPq for financial support (PQ--2 grant, process 
 No. 307628/2022-1).  The work by AYP has been supported by the CNPq project No. 303777/2023-0. GJO also acknowledges financial support from the project i-COOPB23096 (funded by CSIC). The paper is based upon work from COST Actions CosmoVerse CA21136 and CaLISTA CA21109, supported by COST (European Cooperation in Science and Technology).


\end{document}